\documentclass[12pt]{article}
\usepackage{graphicx,amsmath}
\begin{document}


\title{\vskip1cm
Aliasing modes in the lattice Schwinger model 
\vskip.5cm}
\author{Rafael G. Campos and Eduardo S. Tututi\\ 
Facultad de Ciencias F\'{\i}sico--Matem\'aticas, \\
Universidad Michoacana \\
58060 Morelia, Michoac\'an, M\'exico\\[1cm]
{\tt rcampos@umich.mx,   tututi@umich.mx}\\
}
\date{}
\maketitle
{
}\\[1cm]

\begin{center} Abstract \end{center}
We study the Schwinger model on a lattice consisting of zeros of the Hermite polynomials that incorporates a lattice derivative and a discrete Fourier transform with many properties. Such a lattice produces a Klein-Gordon equation for the boson field and the exact value of the mass in the asymptotic limit if the boundaries are not taken into account. On the contrary, if the lattice is considered with boundaries new modes appear due to aliasing effects. In the continuum limit, however, this lattice yields also a Klein-Gordon equation with a reduced mass.
\vfill
\section{Introduction}\label{intro}
Since lattice theory is intended as a nonperturbative way to solve problems in Quantum Field Theory both from analytical and numerical points of view,  it is important to study finite lattice models in order to support consistent numerical calculations as well 
as the belief that an infinite lattice is in fact the continuum limit of a finite one. The Schwinger model has proved to be an excellent theoretical laboratory to study physical properties of fermions on the lattice \cite{Mel00,bod-kov,Sdu04,Sdu05,Sdu06}. Several approaches and derivatives has been used to study the mass spectrum, chirality, charge screening and  the chiral condensate, among others. In fact, in \cite{Mel00} it has shown that different types of fermions derivatives, namely the naive, the Wilson's and the so called modified SLAC derivative work well for the Schwinger model,  yielding the correct boson mass in the continuum limit, but also producing a infinitely heavy boson mode. In the past years, two different differentiation matrices (nonlocal lattice derivatives) have been used to circumvent the restrictions of the no-go Nielsen-Ninomiya theorem \cite{CamTut1,Cam03}. One of them is the Trigonometric derivative defined for $2\pi$-periodic functions \cite{Cam00b}. The other, which can be called Hermite derivative, is suitable for nonperiodic functions decreasing rapidly to zero at infinity. Each derivative has a well-defined discrete Fourier transform. If the number of lattice points is an odd integer, the spectrum of the Trigonometric derivative coincides with that obtained by using the SLAC derivative.\\ 
In this paper we study the lattice Schwinger model using a lattice consisting of zeros of the Hermite polynomials that incorporates a lattice derivative and a discrete Fourier transform. We obtain the correct  boson mass in the asymptotic limit if the boundaries are not taken into account, however, new boson modes appear in the case in which the boundaries of the lattice are considered. This effect is completely due to the aliasing inherent in a finite lattice.\\
Let us start by considering the explicit form of the derivative used in this paper. The $N\times N$ Hermite differentiation matrix $D$ is given by $D=S\tilde{D}S^{-1}$ where
\begin{equation}\label{derher}
\tilde D_{jk}=\begin{cases}
0,&i=j,\\ \noalign{\vskip .5truecm}
\displaystyle {1\over{x_j-x_k}}, &i\not=j,\\
\end{cases} 
\end{equation}
$S$ is a diagonal matrix \cite{CamTut1} which nonzero elements are given by $\exp[-x_k^2/2]H'_N(x_k)$, $k=1,2,\ldots,N$, and $x_k$ is the $k$th zero of the Hermite polynomial $H_N(x)$. The prime means differentiation with respect to the argument. This matrix yields exact derivatives at the lattice points $x_k$ for functions of the form $\exp(-x^2/2)f(x)$, where $f(x)$ is a polynomial of degree less than $N$,
and in general gives approximated values to the derivative of a differentiable square-integrable function, since a function like this can be expanded in a Hermite series. Taking into account that 
\[
H'_N(x_j)=\frac{2N!}{\Gamma[(N+1)/2]}(-1)^{N+j}e^{x_j^2/2}+{\mathcal O}(1/N),
\]
the elements of the matrices $D$ and $\tilde{D}$ coincides asymptotically up to a factor $(-1)^{j+k}$ and therefore, in this limit $D$ becomes a skew-symmetric matrix with a simple structure.
\\
The differentiation matrix is  diagonalized by the discrete Fourier transform $F$. Thus, we have an equation of the form $F^\dagger DF=iP$ where $P$ is a diagonal matrix containing the eigenvalues of $-iD$. The discrete Fourier transform is given by
\[
F_{jk}=\sum_{l=0}^{N-1} (i)^l\varphi_l(x_j)\varphi_l(p_k),\quad P_{kk'}=p_k\delta_{kk'},
\]
where
\[
\varphi_l(\xi)=\sqrt{{{(N-1)!2^{{N}-1-l}}\over{N l!}}}{{H_l(\xi)}\over{H_{N-1}(\xi)}}.
\]
Here, $x_k$ and $p_k$ are roots of $H_N(\xi)$, and if $k$ is fixed, they have the asymptotic form 
\begin{equation}\label{rasinher}
x_k=p_k=\frac{\pi}{\sqrt{2N}}\,k, \quad \text{$k$ integer}.
\end{equation}
In this limit, the discrete Fourier transform  behaves as $F_{jk}=\sqrt{(\Delta x \Delta p/2\pi)}\,\, e^{ix_jp_k}+{\mathcal O}(1/N)$
where $\Delta x=x_{k+1}-x_k=\Delta p=p_{k+1}-p_k=\pi/\sqrt{2N}$. Since both $\Delta x$ and $\Delta p$ are dimensionless and equal, we can write this asymptotic expression in terms of the Riemann measure $\Delta\xi=\Delta x= \Delta p$
\begin{equation}\label{fasi}
F_{jk}=\frac{\Delta\xi}{\sqrt{2\pi}}\,\, e^{ix_jp_k}+{\mathcal O}(1/N).
\end{equation}
\section{The lattice model}\label{latmod}
Our study of the lattice Schwinger model is based on the Heisenberg equations and canonical quantization, therefore, lattice artifacts will be used only for the space variable since time can be considered to remain continuous. In the temporal gauge $A_0=0$, the Hamiltonian of the Schwinger model is given by 
\begin{equation}\label{hamilcont}
H=\int dx \left[\frac{1}{2}E^2(x)+i \psi^\dagger \sigma_z \left(\partial_x+i e  A(x)\right)\psi(x)\right],
\end{equation}
where $A(x)$ stands for the component $A_1(x)$ of the gauge field. To get a discretized version of this Hamiltonian it is necessary to consider the action of any differentiation matrix $D$ on the fermionic variable in more detail. Since
\[
(D\psi^\alpha)_j=\sum_{k=1}^N D_{jk}\psi^\alpha_k,
\]
where $\psi^\alpha$ is the vector whose elements are the values of the $\alpha$-component of the fermionic field $\psi(x)$ at the lattice points and time $t$, the discretized kinetic terms become nonlocal for a finite number of nodes. Thus, for finite $N$ the covariant derivative must be replaced by a matrix differentiation with links. This task can be carried out by noting that under the transformation $x_k\to x_j$ the fermionic field transforms as 
\begin{equation}\label{psitrangen}
\psi(x_k)\to e^{ie\int_{x_j}^{x_k}A(\xi) d\xi}\psi(x_j)=e^{ieb_k}e^{-ieb_j}\psi(x_j),
\end{equation}
where we have written $\int_{x_j}^{x_k}A(\xi) d\xi=b_k-b_j$. Therefore, the covariant derivative $\partial_x+i e  A(x)$ is discretized in our scheme as
\begin{equation}\label{covder}
\nabla=U^\dagger\circ D\circ U,
\end{equation}
where $U$ is the unitary matrix with elements $U_{jk}=\delta_{jk}e^{-ieb_j}$ and $\circ$ denotes the element-wise matrix product.  Since the differentiation matrix (\ref{derher}) become local derivative in the continuum limit, $\nabla$ yields the correct form of the covariant derivative. \\
Thus, our lattice version of (\ref{hamilcont}) is given by
\begin{equation}\label{hamildis}
H_L=\frac{1}{2}E^tE+i \psi^\dagger(\sigma_z \otimes \nabla) \psi,
\end{equation}
where $E$ and $\psi$ are the vectors constructed with the values of the fields $E(x)$ and $\psi(x)$ at the lattice points and the superscript $t$ means transpose.\\
Taking into account the similarity transformation (\ref{covder}), $H_L$ can be written in the form of a free Hamiltonian
\begin{equation}\label{hamildisfree}
H_L=H_\rho+H_\Psi=\frac{1}{2}\rho^t D^{-2}\rho+i \Psi^\dagger(\sigma_z \otimes D) \Psi,
\end{equation}
where $\Psi^\alpha=U\psi^\alpha$ and the discretized form of Gauss' law 
\[
\rho=DE=e\, {\psi^\alpha}^\dagger\circ \psi^\alpha=e\, {\Psi^\alpha}^\dagger\circ \Psi^\alpha
\]
has been used. In order to have a nonsingular matrix $D$, the number of nodes $N$ must be an even integer. The $r$th value of the axial current $j$ is given by 
\[
j_r=-e\, \psi^\dagger_r{^\alpha} \sigma_{\alpha\beta} \psi_r^\beta=-e\,  \Psi^\dagger_r{^\alpha} \sigma_{\alpha\beta} \Psi_r^\beta,
\]
and the fermion fields $\psi$, $\psi^\dagger$ satisfy the equal-time anticommutation relations
\begin{equation}\label{anticomrel}
\{\psi^\dagger_r{^\alpha},\psi^\beta_s\}=\delta_{rs}\delta_{\alpha\beta}.
\end{equation}
Note that in order to make the connection to the continuum case, $H_L$ should be multiplied by $\Delta\xi$ to obtain the Riemann sum of (\ref{hamilcont}), whereas (\ref{anticomrel}) should be divided by $\Delta\xi$ to obtain a delta function.
\section{An equation for the charge density}\label{eqch}
To find an equation for the bosonic variable $\rho$ on the lattice we begin by considering the lattice version of $\partial_0 \rho=[\rho,H]/i$, which can be written as $(\partial_0 \rho)_r=[\rho_r,H_L]/i$. The error in the derivative of a field produced by the use of a differentiation matrix is negligible if the number $N$ of points is great, therefore, we assume a lattice with a sufficiently great number of nodes. There are, however, two possibilities to consider in this asymptotic limit. One is what can be called an infinite lattice, i.e., a lattice with no boundaries and therefore, no border effects. The other is a finite lattice with definite boundaries. The following first calculations are common for both cases since they only use the fact that $N$ is sufficiently great.\\
By using (\ref{anticomrel}) and the skew-symmetry of $D$ we obtain the continuity equation
\begin{eqnarray}\label{eqcont}
(\partial_0\rho)_r&=&[\rho_r,H_L]/i=e\left(\Psi^\dagger_r{^\alpha}\sigma_{\alpha\beta}D_{rs}\Psi^\beta_s-\Psi^\dagger_s{^\alpha}\sigma_{\alpha\beta}D_{sr}\Psi^\beta_r\right)\nonumber\\
&=&e\left(\Psi^\dagger_r{^\alpha}\sigma_{\alpha\beta}(D\Psi^\beta)_r+(D\Psi^\dagger{^\alpha})_r\sigma_{\alpha\beta}\Psi^\beta_r\right)\\
&=&e [D( \Psi^\dagger{^\alpha} \sigma_{\alpha\beta} \Psi^\beta)]_r=-(Dj)_r.\nonumber
\end{eqnarray}
Now, let us compute 
\begin{equation}\label{segderro}
(\partial^2_0 \rho)_r=-[(Dj)_r,H_L]/i=-[(Dj)_r,H_\rho]/i-[(Dj)_r,H_\Psi]/i.
\end{equation}
Proceeding as before, we find
\begin{eqnarray}\label{segderrodos}
[(Dj)_r,H_\Psi]/i&=&-e\big(\Psi^\dagger_r[(1_2\otimes D^2)\Psi]_r+2[(1_2\otimes D)\Psi]^\dagger_r[(1_2\otimes D)\Psi]_r
+[(1_2\otimes D^2)\Psi]^\dagger_r\Psi_r\big)\nonumber\\
&=&-(D^2\rho)_r,
\end{eqnarray}
where $1_2$ is the identity matrix of dimension 2. Therefore, (\ref{segderro}) becomes
\begin{equation}\label{kleingoru}
(\partial^2_0 \rho)_r-(D^2\rho)_r=-\frac{1}{2i}[(Dj)_r,\rho^t D^{-2}\rho].
\end{equation}
To obtain the asymptotic form of the mass term it is necessary to compute the Schwinger commutator $[j_j,\rho_{j'}]$. To this end we compute  the commutator of Fourier transforms $[\tilde{j}_k,\tilde{\rho}_{k'}]$. First, we expand $\Psi^{\alpha}_q$ in its Fourier components
\[
\Psi^{\alpha}_q=\sum_k[ a_ku^{\alpha}_k F_{qk} + b^\dagger_k{v^\dagger}^{\alpha}_k F^\dagger_{kq}],
\]
where $F$ stands for the discrete Fourier transform (\ref{fasi}). Thus, the Fourier representation of 
$\rho_q=e\Psi^\dagger_q \Psi_q$ is
\[
\rho_q=e\sum_{k,k'}[ a^\dagger_ka_{k'}F^\dagger_{kq} F_{q{k'}} + b_kb^\dagger_{k'} F_{qk}F^\dagger_{{k'}q}],
\]
where we have used the orthogonal property of the spinors. Therefore, the Fourier transform $\tilde{\rho}_k=\sum_qF^\dagger_{kq}\rho_q$ becomes
\begin{equation}\label{roprdfs}
\tilde{\rho}_k=e\sum_{q,k',k''}[ a^\dagger_{k''}a_{k'}F^\dagger_{k''q}F_{q{k'}}F^\dagger_{kq} +
 b_{k''}b^\dagger_{k'}F_{q{k''}}F^\dagger_{{k'}q}F^\dagger_{{k}q}].
\end{equation}
Note that the sum of products of elements of the discrete Fourier transform $F$ involved in this equation are proportional to the generic asymptotic form
\begin{equation}\label{prodfs}
\frac{1}{2\pi}\sum_q e^{i x_q P}\Delta\xi,
\end{equation}
where $P$ stands for a sum of three lattice momenta, for instance, $P=p_{k'}-p_k-p_{k''}$. It is worth to notice that a sum like (\ref{prodfs}) is exactly a Kronecker delta if $P$ is the sum of {\it two} lattice momenta (since $F$ is unitary). Therefore, (\ref{prodfs})
becomes an exact delta whenever the sum of two of the three momenta summing up $P$ is again an allowed lattice momentum. But this is not always the case, even for uniformly spaced momenta. However, we can consider an infinite lattice, i.e., a lattice with a very great number of nodes with no boundaries and extending to infinity. Since there is a momentum for each lattice site, the lattice momenta do not have neither a maximum nor a minimum and therefore, the sum of two evenly spaced momenta is again an allowed momentum, yielding that
\begin{equation}\label{sumdelp}
\frac{\Delta\xi}{2\pi}\sum_q e^{i x_q P}\Delta \xi\to \delta(P)\Delta\xi.
\end{equation}
In this case, (\ref{roprdfs}) becomes
\begin{equation}\label{rhouno}
\tilde{\rho}_k=e\frac{\Delta \xi}{\sqrt{2\pi}}\,\,\sum_{k'}[ a^\dagger_{k'} a_{k'+k}+b_{k'} b^\dagger_{k'-k}].
\end{equation}
A similar computation can be done for $\tilde{j}_k$. To obtain the value of $[\tilde{j}_k,\tilde{\rho}_{k'}]$, it is necessary to write
this commutator in terms of operators that are normal-ordered with respect to the vacuum state filled with particles of any negative energy and antiparticles of any positive energy, defined as 
\[
\vert \text{vac}\rangle=\prod_{k<0}a^\dagger_k\prod_{k>0}b^\dagger_k\vert 0 \rangle.
\] 
Thus, after the use of the anticommutation relations 
\begin{equation}\label{anticomopcd}
\{a^\dagger_k,a_{k'}\}=\delta_{kk'}, \quad \{b^\dagger_k,b_{k'}\}=\delta_{kk'},
\end{equation}
the normal-ordered commutator becomes\footnote{Again, in order to make the connection to the continuum case, (\ref{anticomopcd}) should be divided by $\Delta \xi$ to obtain a delta function.}
\[
[\tilde{j}_k,\tilde{\rho}_{k'}]=e^2\frac{\Delta \xi}{2\pi}\sum_{m}(a^\dagger_m a_{m+k+k'}-a^\dagger_{m-k} a_{m+k'}+b^\dagger_{m-k-k'}b_{m}-b^\dagger_{m-k'}b_{m+k}  ),
\]
and, as it can be seen, it annihilates the vacuum state unless $k'=-k$. Therefore, 
\[
[\tilde{j}_k,\tilde{\rho}_{k'}]=e^2\delta_{k',-k}\frac{\Delta \xi}{2\pi}\sum_{m}(a^\dagger_m a_m-a^\dagger_{m-k} a_{m-k}+b^\dagger_m b_{m}-b^\dagger_{m+k}b_{m+k}  )
\]
and the action of this commutator on $\vert \text{vac}\rangle$ is given by 
\[
[\tilde{j}_k,\tilde{\rho}_{k'}]\vert \text{vac}\rangle=-\,\frac{e^2}{\pi}\,\Delta \xi\,\,k\,\delta_{k',-k}\vert \text{vac}\rangle.
\]
Thus, the inverse transform of $[\tilde{j}_k,\tilde{\rho}_{k'}]$, is
\begin{equation}\label{conmjjp}
[j_j,\rho_{j'}]=-\,\frac{e^2}{\pi}\,\Delta \xi\,\sum_k k F_{jk}F_{j',-k}=-\,\frac{e^2}{\pi}\,\Delta \xi\,\sum_k k F_{jk}F^\dagger_{kj'}.
\end{equation}
The product $\Delta \xi\,\, k F_{jk}$ is the asymptotic form of $p_k F_{jk}=-i\sum_l D_{jl}F_{lk}$, as it can be seen from (\ref{rasinher}). By using the fact that $FF^\dagger=1$, we obtain finally the Schwinger commutator on an infinite lattice
\begin{equation}\label{schwcomminf}
[j_j,\rho_{j'}]=i\frac{e^2}{\pi}\,D_{jj'}.
\end{equation}
Therefore, the right-hand side of (\ref{kleingoru}) becomes the correct value of the mass term
\begin{equation}\label{massterm}
m^2\rho_r=\frac{1}{2i}[(Dj)_r,\rho^t D^{-2}\rho]=\frac{e^2}{\pi}\rho_r.
\end{equation}
Let us consider now the case of a finite lattice, i.e., a lattice with a great number of nodes and definite boundaries. Since momenta and sites have same numerical values, the allowed momenta lie in the same interval as the sites do. Thus, in the case of a finite lattice, the sum of evenly spaced lattice momenta may lie outside the interval of momenta and the exponential in (\ref{prodfs}) produces aliasing in the momenta domain\footnote{Aliasing is an effect associated with discretization of periodic continuous systems. In a video, for instance, the wheels of a car appear to be static or to rotate in reverse. This effect is due to aliasing of the wheel rotation frequency (see for example \cite{Gas99})}. To explain in more detail the aliasing effect let us use (\ref{rasinher}) to rewrite  (\ref{prodfs}) as
\begin{equation}\label{prodfsdos}
\frac{1}{2\pi}\sum_q \exp (i \frac{\pi^2}{2N} qK)\Delta\xi=\frac{1}{2\pi}\sum_q \exp [i2\pi q (\pi \kappa/4)]\Delta\xi,
\end{equation}
where $\kappa=K/N$, $q$ is a nonzero integer of $[-N/2,N/2]$ and $K$ is a sum of integers of $[-N/2,N/2]$. In the asymptotic limit $\kappa$ becomes dense in the real line and two momenta $p_k$ and $p_{k'}$ are indistinguishable for the model if their indexes, $\kappa$ and $\kappa'$, are congruent module $4/\pi$, i.e., if they yield the same sum (\ref{prodfsdos}). In other words, a momentum $p_k$ is distinguishable if its index $\kappa$ lies inbound in the interval $[-2/\pi,2/\pi]$. Note that this is the interval of periodicity of the exponential function of (\ref{prodfsdos}). Thus, the sum of two lattice momenta (whose indexes are in $[-1/2,1/2]$) can be a lattice momentum again with $\kappa$ lying in $[-1/2,1/2]$, or a totally new mode with $\kappa\in [-2/\pi,-1/2]\cup[1/2,2/\pi]$, or an aliased and repeated mode with $\kappa\in [-1,1/2-4/\pi]\cup[4/\pi-1/2,1]$. The interval $[4/\pi-1/2,1]$ is aliased with $[-1/2,1-4/\pi]$ and $[-1,1/2-4/\pi]$ with $[4/\pi-1,1/2]$, such as depicted in Figure 1. 
\vskip1cm
\hbox to \textwidth{\hfill\scalebox{0.8}{\includegraphics{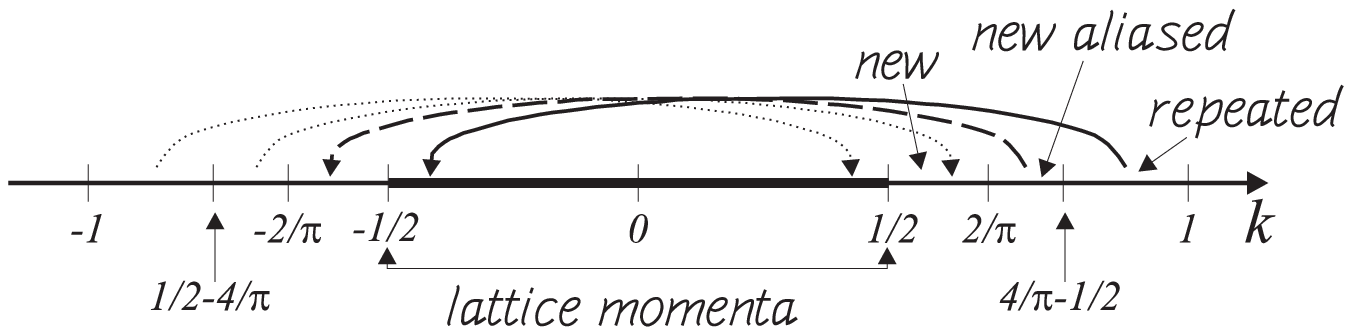}}\hfill}
\begin{center}
\begin{minipage}{10cm}
{\small 
Figure 1: Aliased modes appearing in a finite lattice.
}
\end{minipage}
\end{center}
Since  $N$ is sufficiently great, (\ref{sumdelp}) still holds in the form
\begin{equation}\label{sumdelptil}
\frac{\Delta\xi}{2\pi}\sum_q e^{i x_q (p_{k'}-p_k-p_{k''})}\Delta \xi\to \delta(p_{k'}-(\overline{p_k+p_{k''}}))\Delta\xi.
\end{equation}
where $\overline{p_k+p_{k''}}$ stands for the value (aliased or not) corresponding to the sum $p_k+p_{k''}$. Note that the aliasing does not occur if one of the two momenta has an index $\vert \kappa \vert < 4/\pi-1\approx 0.27$, i.e, $\kappa\in[1-4/\pi,4/\pi-1]$. For other values of $\kappa$, the aliasing effect changes the charge density (\ref{rhouno}) in the momenta space  to 
\begin{eqnarray*}\label{rhoalias}
\tilde{\rho}_{\kappa'}&=&e\frac{\Delta \xi}{\sqrt{2\pi}}\,\Bigl(\sum_{\eta\in[-\frac{1}{2},\frac{4}{\pi}-\frac{1}{2}-{\kappa'}]} a^\dagger_{\eta} a_{\eta+{\kappa'}}+\sum_{\eta\in[\frac{4}{\pi}-\frac{1}{2}-{\kappa'},\frac{1}{2}]} a^\dagger_{\eta} a_{\eta+{\kappa'}-\frac{4}{\pi}}\nonumber\\&+&
\sum_{\eta\in[-\frac{1}{2},\frac{1}{2}-\frac{4}{\pi}+{\kappa'}]} b_{\eta} b^\dagger_{\eta-{\kappa'}+\frac{4}{\pi}}+
\sum_{\eta\in[\frac{1}{2}-\frac{4}{\pi}+{\kappa'},\frac{1}{2}]} b_{\eta} b^\dagger_{\eta-{\kappa'}}\Bigr),\quad {\kappa'}>4/\pi-1,
\end{eqnarray*}
whereas the axial current becomes
\begin{eqnarray*}\label{jotalias}
\tilde{j}_\kappa&=&e\frac{\Delta \xi}{\sqrt{2\pi}}\,\Bigl(-\sum_{\eta\in[-\frac{4}{\pi}+\frac{1}{2}-\kappa,\frac{1}{2}]} a^\dagger_{\eta} a_{\eta+\kappa}-\sum_{\eta\in[-\frac{1}{2},-\frac{4}{\pi}+\frac{1}{2}-\kappa]} a^\dagger_{\eta} a_{\eta+\kappa+\frac{4}{\pi}}\nonumber
\\&+&
\sum_{\eta\in[\frac{4}{\pi}-\frac{1}{2}+\kappa,\frac{1}{2}]} b_{\eta} b^\dagger_{\eta-\kappa-\frac{4}{\pi}}+
\sum_{\eta\in[-\frac{1}{2},\frac{4}{\pi}-\frac{1}{2}+\kappa]} b_{\eta} b^\dagger_{\eta-\kappa}\Bigr),\quad \kappa<1-4/\pi.
\end{eqnarray*}
As before, the normal-ordered Schwinger commutator $[\tilde{j}_k,\tilde{\rho}_{k'}]$ annihilates the vacuum state unless  $\kappa'=-\kappa$ (or $k'=-k$). In this case the action on $\vert \text{vac}\rangle$ is
\begin{equation}\label{actconmkalias}
[\tilde{j}_k,\tilde{\rho}_{k'}]\vert \text{vac}\rangle=\frac{e^2}{\pi}\,\Delta \xi\,\,\delta_{k',-k}
\begin{cases}-\,k\,\vert \text{vac}\rangle,& \vert\kappa\vert, \vert\kappa'\vert\le\frac{4}{\pi}-1,\\
N(1-\frac{4}{\pi})\vert \text{vac}\rangle,& \kappa<1-\frac{4}{\pi},\\
N(\frac{4}{\pi}-1)\vert \text{vac}\rangle,&\kappa>\frac{4}{\pi}-1,
\end{cases}
\end{equation}
and now (\ref{conmjjp}) becomes
\begin{eqnarray*}
[j_j,\rho_{j'}]&=&-\,\frac{e^2}{\pi}\,\Delta \xi\,\sum_{\vert k\vert\le (4/\pi-1)N} k F_{jk}F^\dagger_{kj'}\\
&+&i\frac{e^2}{\pi^2}\,\Delta \xi^2\,(4/\pi-1)N\sum_{k> (4/\pi-1)N}^{N/2}\sin p_k(x_j-x_{j'})\Delta \xi.
\end{eqnarray*}
The first sum of the right-hand side corresponds to the values of the momentum index for which aliasing does not occur and it approaches to (\ref{schwcomminf}) in the asymptotic limit. The second sum can be approximated by an integral to yield
\begin{equation}\label{consch}
[j_j,\rho_{j'}]=i\frac{e^2}{\pi}\,D_{jj'}+i\frac{e^2}{2}(4/\pi-1) (-1)^{j+{j'}}S(y_{jj'})\,D_{jj'},
\end{equation}
where $S(y)=[\cos(4/\pi-1)y-\cos y/2]$, $y_{jj'}=\pi\sqrt{N/2}\,(x_j-x_{j'})$ and we have used the fact that 
$1/(x_j-x_{k})=(-1)^{j+k}D_{jk}$. The substitution of this commutator in the mass term of (\ref{kleingoru}) yields two terms, one of them 
corresponds to the first term of the right-hand side of (\ref{consch}) and it gives the correct mass $e^2/\pi$. Due to $S(y_{jj'})$, the remaining term yields a mixing of modes $\rho_{r'}$ at locations different from $r$ for a finite $N$ [cf. Eq. (\ref{massterm})]. However, since $\vert S(y_{jj'})\vert\le 2$, the behavior of $S(y_{jj'})\,D_{jj'}$ is dominated by $D_{jj'}$ and in the continuum limit, i.e., in the limit $N\to\infty$, $S(y_{jj'})\,D_{jj'}$ becomes essentially $D_{jj'}$ due to the singular behavior of the matrix differentiation \cite{Cam03}. Thus, in such a limit, the main contribution of $S(y_{jj'})$ is given by its first maximum (see Figure 2) located at $y_{jj'}\approx \pm 5.14$, i.e., at $x_j-x_{j'}\approx \pm\pi/\sqrt{2N}=\pm\Delta\xi$, or equivalently, at $j'=j\pm 1$. 
\vskip1cm
\hbox to \textwidth{\hfill\scalebox{0.8}{\includegraphics{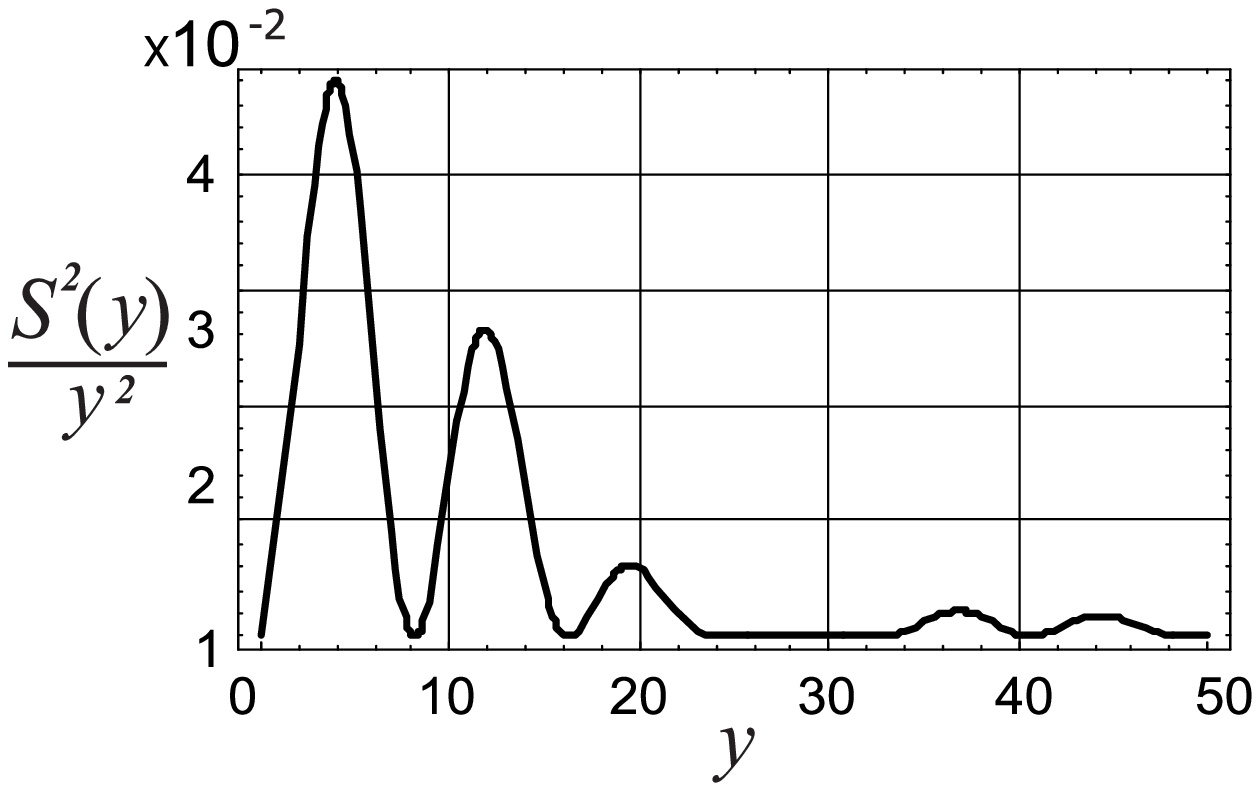}}\hfill}
\begin{center}
\begin{minipage}{10cm}
{\small 
Figure 2: Plot of the function $S^2(y)/y^2$
}
\end{minipage}
\end{center}
\vskip.5cm
Since $\vert S(5.14)\vert\approx 1$, we have that $\vert S(y_{jj'})\vert \,D_{jj'}\approx D_{jj'}$, $N\to\infty$,
and therefore, in the continuum limit (\ref{consch}) behaves as
\begin{equation}\label{conschcl}
[j_j,\rho_{j'}]=i\frac{e^2}{\pi}\,D_{jj'}-i\frac{e^2}{2}(4/\pi-1) \,D_{jj'},
\end{equation}
yielding
\begin{equation}\label{masstermfin}
m^2=(\frac{1}{2}-\frac{1}{\pi})e^2=0.18e^2
\end{equation}
for the square of the mass of the Schwinger model on a lattice with boundaries.\\
It is important to stress that aliasing is not just a feature of the discrete Fourier transform used in this paper and that 
aliasing effects should appear in lattice field theory whenever a {\it finite} lattice and the usual discrete Fourier transform
are used to compute a quantity which depends on the sum of more than two momenta, as in Eq (\ref{sumdelptil}). The
equivalent form of this equation (using the usual Fourier transform) is 
\[
\sum_{q=1}^N e^{\displaystyle{i2\pi\frac{q}{N}}(k'-k-k'')},
\]
where $k,k',k''=1,2,\ldots,N$, and we have omitted the normalization factors. In this case, we have again aliasing effects since
the sum $k+k''$ may be an integer greater than $N$ and therefore may lie outside the finite lattice. Quantities involving more
than two Fourier transforms appear for example, when computing vertex functions.\\
\section{Concluding remarks}
We have introduced a nonperiodic lattice that incorporates a lattice derivative and a discrete Fourier transform with useful properties. Such a lattice has potential use in the study of quantum-field problems. In the case of the Schwinger model, the use of this nonperiodic lattice yields the correct bosonic mass if the lattice is considered unbounded. On the other hand, if the lattice has boundaries, a great number of new modes appears due to the inherent aliasing in the finite Fourier transform. In fact we have shown that in this case, the boson acquires a smaller mass. Since aliasing is an effect present in the usual discrete Fourier transform, the appearance of new modes or aliasing effects should be expected in any lattice approach where the standard discrete Fourier transform is used. Such effects, mainly due to the boundedness of the lattice, indicate that the results obtained by numerical calculations using (necessarily) finite lattice models of quantum-field problems defined on unbounded manifolds may not be completely correct if the aliasing effects are not taken into account.

\end{document}